\documentclass[aps,twocolumn,showpacs, showkeys,nofootinbib,floatfix]{revtex4-1}

\usepackage{amssymb}
\usepackage{amsmath}
\usepackage{graphicx}
\usepackage{hyperref}
\usepackage{longtable}

\begin{document}

\title{Unified Dark Fluid with Constant Adiabatic Sound Speed: Including Entropic Perturbations}

\author{Lixin Xu}
\email{lxxu@dlut.edu.cn}

\affiliation{Institute of Theoretical Physics, School of Physics \&
Optoelectronic Technology, Dalian University of Technology, Dalian,
116024, P. R. China}

\affiliation{College of Advanced Science \& Technology, 
Dalian University of Technology, Dalian, 116024, P. R. China}

\begin{abstract}

In this paper, we continue to study a unified dark fluid model with a constant adiabatic sound speed but with the entropic perturbations. When the entropic perturbations are included, an effective sound speed, which reduces to the adiabatic sound speed when the entropic perturbations are zero, has to be specified as an additional free model parameter. Due to the relations between the adiabatic sound speed and equations of state (EoS) $c^2_{s,ad}(a)=w(a)-d\ln(1+w(a))/3 d\ln a$, the equation of state can be determined up to an integration constant in principle when an adiabatic sound speed is given. Then there are two degrees of freedom to describe the linear perturbations for a fluid. Its micro-scale properties are characterized by its EoS or adiabatic sound speed and an effective sound speed. We take the effective sound speed and adiabatic sound speed as free model parameters and then use the currently available cosmic observational data sets, which include type Ia supernova Union 2.1, baryon acoustic oscillation and WMAP $7$-year data of cosmic background radiation, to constrain the possible entropic perturbations and the adiabatic sound speed via the Markov Chain Monte Carlo method. The results show that the cosmic observations favor a small effective sound speed $c^2_{s,eff}=0.00155_{-    0.00155}^{+    0.000319}$ in $1\sigma$ region. 

\end{abstract}



\maketitle

\section{Introduction}

In the last few years, the unified dark fluid models \cite{ref:darkdeneracy,ref:Bruni,ref:darkdeneracyxu,GCG,GCG-action,GCGpapers,GCGdecomp,GCGxu,ref:xuNUDF,ref:alphcdm,ref:cs0,ref:csvar} were investigated as a possible explanation to an accelerated expansion phase of our Universe \cite{ref:Riess98,ref:Perlmuter99}. These models are inspired by the facts that above $96\%$ of the energy content in the Universe is made of unknown dark component. These unified dark fluid models include the popular generalized Chaplygin gas (gCg) model \cite{GCG,GCG-action,GCGpapers,GCGdecomp,GCGxu} as a sample which is a generalization of the Chaplygin gas (Cg) model or a coined model from the $\Lambda$CDM model \cite{ref:xuNUDF}. Actually, the EoS' of these unified dark fluid are specified in different models. Then their linear perturbation evolutions are determined via the relations $c^2_{s,ad}(a)=w(a)-d\ln(1+w(a))/3 d\ln a$ between the adiabatic sound speed and its EoS of UDF when the entropic perturbations are zero. On the contrary, when the adiabatic sound speed is preposed, the EoS can also be determined up to an integration constant in principle. Based on this point, the so-called $\Lambda\alpha$CDM or CASS model was proposed \cite{ref:Bruni,ref:darkdeneracyxu,ref:alphcdm} where a constant adiabatic sound speed $c^2_{s,ad}=\alpha$ was assumed. The case of $\alpha=0$ was discussed in Ref. \cite{ref:cs0}. And the time variable sound speed cases were also discussed in \cite{ref:csvar}. As pointed out by the authors \cite{alphanegative}, when $\alpha$ is negative, this model can be seen as the attractor solution of quintessence scalar field dynamics. Also, it can be treated as k-essence scalar field \cite{ref:darkdeneracyxu}. Interestingly, it can avoid the so-called averaging problem \cite{ref:stability} when the perturbations become nonlinear \cite{ref:darkdeneracyxu}. So, in this paper, we are going to investigate this unified dark fluid model.      

However when the entropic perturbations are included the adiabatic sound speed is not enough to characterize the micro-scale properties, one has to introduce the effective sound speed as a specified model parameter as an addition to its EoS. For a generalized dark matter, the effective sound speed was defined in Ref. \cite{ref:Hu98}. The effective sound speed is reduced to the adiabatic sound speed when the entropic perturbations vanish.  

The model with a constant adiabatic sound speed and entropic perturbations was investigated in Ref. \cite{ref:alphaentropy}, where the limited cases of effective sound speed $c^2_{eff}=\alpha,0,1$ were discussed. However, the effective sound speed should be a free model parameter to be determined by the cosmic observations instead of being fixed to a special value by hand. So, in this paper, we will consider a more general case where the effective sound speed $c^2_{s,eff}$ is taken as a free model parameter in the range of $[0,1]$. And we will use the currently available cosmic observational data sets, which include type Ia supernova Union 2.1, baryon acoustic oscillation and WMAP $7$-year CMB, to determine the model parameter space via the Markov Chain Monte Carlo (MCMC) method.

This paper is structured as follows. At first, in section \ref{sec:DG}, we give a very brief review of the unified dark fluid model with a constant adiabatic sound speed (CASS) $c^2_{s,ad}=\alpha$. In this section, the energy density, EoS and the perturbation equations are shown. In section \ref{sec:method}, by using the Markov Chain Monte Carlo (MCMC) method with currently available cosmic observational data sets which include type Ia supernova Union 2.1, baryon acoustic oscillation and WMAP $7$-year CMB, we show the model parameter space. A summary is presented in section \ref{ref:conclusion}.

\section{A brief review of CASS model}   \label{sec:DG}  

In this section, we will give a very brief review of CASS model which has constant adiabatic sound speed $c^2_s=\alpha$, for the details please see the Ref. \cite{ref:darkdeneracyxu}. The energy density and equation of state (EoS) of this UDF are given in the following forms
\begin{eqnarray}
\rho_d&=&\rho_{d0}\left[(1-B_s)+B_s a^{-3(1+\alpha)}\right],\label{eq:rhod}\\
w_d&=&\alpha-\frac{(1+\alpha)(1-B_s)}{(1-B_s)+B_s a^{-3(1+\alpha)}},
\end{eqnarray}
where $B_s$ in the range $0\le B_s\le 1$ and $\alpha$ are model parameters. In the pure adiabatic perturbation case, the value of $\alpha$ should be fixed in the range of $[0,1]$. When the entropic perturbations are included, the adiabatic sound speed can be negative \cite{ref:Hu98}. So, in this paper, we assume it is in the range $[-1,1]$. Of course it will be determined by the cosmic observations.

Considering the perturbation in the synchronous gauge, the perturbed metric reads 
\begin{equation}
ds^2=a^2(\tau)\left[-d\tau^2+(\delta_{ij}+h_{ij}(\bold{x},\tau)dx^i dx^j)\right],
\end{equation} 
where $\tau$ is the conformal time and $h_{ij}$ is the metric perturbation. From the conservation of energy-momentum tensor $T^{\mu}_{\nu;\mu}=0$, one has the perturbation equations of density contrast and velocity divergence for dark fluid in the synchronous gauge
\begin{eqnarray}
\dot{\delta}_d&=&-(1+w_d)(\theta_d+\frac{\dot{h}}{2})-3\mathcal{H}(\frac{\delta p_{d}}{\delta \rho_{d}}-w_d)\delta_d,\label{eq:continue}\\
\dot{\theta}_d&=&-\mathcal{H}(1-3c^2_{s,ad})+\frac{\delta p_{d}/\delta \rho_{d}}{1+w_d}k^{2}\delta_d-k^{2}\sigma_d\label{eq:euler}
\end{eqnarray}
following the notations of Ma and Bertschinger \cite{ref:MB}, where the definition of the adiabatic sound speed
\begin{equation}
c^2_{s,ad}=\frac{\dot{p}_d}{\dot{\rho}_d}=w_d-\frac{\dot{w}_d}{3\mathcal{H}(1+w_d)}
\end{equation}
 is used. For the gauge ready formalism about the perturbation theory, please see \cite{ref:Hwang}. For a pure barotropic fluid, it has an imaginary adiabatic sound speed which causes instability of the perturbations when its EoS is negative, for example the $w=constant$ quintessence dark energy model. The way to overcome this problem is to allow an entropy perturbation and to assume a positive or null effective speed of sound. Following the formalism for a generalized dark matter \cite{ref:Hu98}, one can separate out the non adiabatic stress or entropy perturbation for the UDF
 \begin{equation}
 p_d\Gamma_d=\delta p_d-c^2_{s,ad}\delta \rho_d, \label{eq:entropyper}
\end{equation} 
 which is gauge independent. In the rest frame of UDF by introducing the effective speed of sound $c^2_{s,eff}$, the entropy perturbation is specified as
 \begin{equation}
 w_d\Gamma_d=(c^2_{s,eff}-c^2_{s,ad})\delta^{rest}_{d}.\label{eq:restframe}
 \end{equation}
 The gauge transformation into an arbitrary gauge 
 \begin{equation}
 \delta^{rest}_{d}=\delta_d+3\mathcal{H}(1+w_d)\frac{\theta_d}{k^2}\label{eq:gaugetrans}
 \end{equation}  
 gives a gauge-invariant form for the entropy perturbations. By using the Eqs (\ref{eq:entropyper},) (\ref{eq:restframe}) and (\ref{eq:gaugetrans}), one can recast Eqs. (\ref{eq:continue}), and (\ref{eq:euler}) into 
 \begin{widetext}
 \begin{eqnarray}
 \dot{\delta}_d&=&-(1+w_d)(\theta_d+\frac{\dot{h}}{2})+\frac{\dot{w}_d}{1+w_d}\delta_d-3\mathcal{H}(c^2_{s,eff}-c^2_{s,ad})\left[\delta_d+3\mathcal{H}(1+w_d)\frac{\theta_d}{k^2}\right]\\
\dot{\theta}_d
&=&-\mathcal{H}(1-3c^2_{s,eff})\theta_d+\frac{c^2_{s,eff}}{1+w_d}k^2\delta_d-k^2\sigma_d
 \end{eqnarray}
\end{widetext}
please see also in \cite{ref:alphaentropy} and \cite{ref:WangXu} for the conformal gauge. The above two equations reduce to the corresponding continuity and Euler equations as shown in our previous work \cite{ref:darkdeneracyxu} when the entropy perturbation vanishes. For the dark fluid in this paper, we assume the shear perturbation $\sigma_d=0$. In our calculations, the adiabatic initial conditions will be taken. To perform the numerical calculation, we modified the publicly available {\bf cosmoMC} package \cite{ref:MCMC} to include the dark fluid perturbation in the {\bf CAMB} \cite{ref:CAMB} code which is used to calculate the theoretical CMB power spectrum. We added new parameters and modified the perturbations equations for the dark energy in the {\bf CAMB} code.

\section{Constraint method and results}\label{sec:method}

\subsection{Implications on CMB anisotropy for model parameter $c^2_{s,eff}$}

In our previous paper \cite{ref:darkdeneracyxu}, for the only adiabatic perturbation case, we have discussed the implication on CMB anisotropic power spectra of the model parameter $\alpha$ and $B_s$, for the details please see \cite{ref:darkdeneracyxu}. When we fix the values of the adiabatic and effective sound speeds, the other relevant model parameters have the same effects to the CMB power spectra as shown in our previous paper \cite{ref:darkdeneracyxu}. So, in this paper, we focus on the model parameter $c^2_{s,eff}$. To illustrate how the CMB temperature anisotropies are characterized by different values of $c^2_{s,eff}$, we choose different values of $c^2_{s,eff}$ in the range $[0,1]$ with the other relevant cosmological models fixed to the mean values obtained in \cite{ref:darkdeneracyxu}. 

\begin{center}
\begin{figure}[htb]
\includegraphics[width=9.cm]{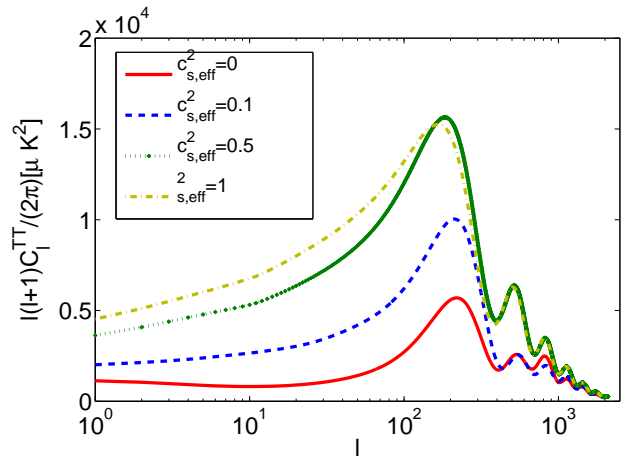}
\caption{The effects on CMB temperature power spectra of model parameter $c^2_{s,eff}$. The solid red, blue dashed, green dotted and yellow dash-dotted lines are for $c^2_{s,eff}=0,0.1,0.5,1$ respectively, where the other relevant cosmological parameters are fixed to their mean values obtained in Ref. \cite{ref:darkdeneracyxu}. The CMB power spectrum shows that small values of $c^2_{s,eff}$ are favored.}\label{fig:cmb}
\end{figure}
\end{center}

For different values of the effective sound speed in the range $[0,1]$, we plotted their effects on the CMB anisotropic power spectra in Figure \ref{fig:cmb}. As shown in this figure, CMB power spectra favor small values of $c^2_{s,eff}$. For large values of $c^2_{s,eff}$, the gravitational potential decays fastly due to pressure support of the UDF fluctuations during UDF domination. The effect monotonically decrease as $c^2_{s,eff}$ decrease to zero which was analyzed in Ref. \cite{ref:Hu98}. It is clear that CMB power spectra favor small values of the effective sound speed. We expect the CMB power spectra can give a tight constraint to the model parameter $c^2_{s,eff}$.   

\subsection{Method and data points}

To constrain the model parameter space, we use the Markov Chain Monte Carlo (MCMC) method. We adopted the following $8$-dimensional parameter space 
\begin{equation}
P\equiv\{\omega_{b},\Theta_{S},\tau, \alpha, B_{s}, c^2_{s,eff}, n_{s},\log[10^{10}A_{s}]\}.
\end{equation}
The priors of the model parameters are shown in Table \ref{tab:priors}. We adopted $k_{s0}=0.05\text{Mpc}^{-1}$ as the pivot scale of the initial scalar power spectrum.

\begin{table}[tbh]
\begin{center}
\begin{tabular}{cc}
\hline\hline Model Prameters & Priors \\ \hline
$\Omega_b h^2$ & $[0.005,1]$ \\
$\Theta_S$ & $[0.5,10]$ \\
$\tau$ & $[0.01,0.8]$ \\
$\alpha$ & $[-1,1]$ \\
$B_s$ & $[0,1]$ \\
$c^2_{s,eff}$ & $[0,1]$\\
$n_s$ & $ [0.5,1.5]$ \\
$\log[10^{10} A_s]$ & $[2.7,4]$ \\
Age & $10\text{Gyr}<t_{0}<\text{20Gyr}$\\
$\omega_{b}$ & $0.022\pm0.002$ \cite{ref:bbn}\\
$H_{0}$ & $74.2\pm3.6\text{kms}^{-1}\text{Mpc}^{-1}$ \cite{ref:hubble}\\
\hline\hline
\end{tabular}
\caption{The priors for the cosmological model parameters and other priors. The pivot scale of the initial scalar power spectrum $k_{s0}=0.05\text{Mpc}^{-1}$ is used in this paper.}\label{tab:priors}
\end{center}
\end{table}

The total likelihood $\mathcal{L} \propto e^{-\chi^{2}/2}$ should be calculated to get the model parameter space, where $\chi^{2}$ is given as
\begin{equation}
\chi^{2}=\chi^{2}_{CMB}+\chi^{2}_{BAO}+\chi^{2}_{SN}.
\end{equation}
For CMB data set, the temperature power spectrum from WMAP $7$-year data \cite{ref:wmap7} are employed. For the BAO information, the SDSS data points \cite{ref:BAO} are used. For SN Ia, we use the $580$ Union2.1 data sets with systematic errors \cite{ref:Union21}. For the detailed description, please see Refs. \cite{ref:darkdeneracyxu,ref:Xu}.

\subsection{Fitting Results and discussion}

We ran $8$ chains in parallel on the {\it Computational Cluster for Cosmos} (3C) and checked the convergence ($R-1$ is of the order $0.01$). The obtained results are shown in Table \ref{tab:results} and Figure \ref{fig:contour}. 
\begin{center}
\begin{table}
\begin{tabular}{lllll}
\hline\hline Prameters&Mean Values & $1\sigma$ errors & $2\sigma$ errors & $3\sigma$ errors \\ \hline
$\Omega_b h^2$ & $0.0228$ & $_{- 0.000628}^{+0.000628}$ & $_{-0.00120}^{+    0.00128}$ & $_{-0.00176}^{+    0.0020315}$ \\
$\Theta_S$ & $1.0495 $ & $_{-0.00271}^{+0.00274} $ & $_{-0.00522}^{+0.00553}$ & $_{-0.00780}^{+0.00844}$ \\
$\tau$ & $0.0949$ & $_{-0.00837}^{+0.00690}$ & $ _{-0.0253}^{+0.0282}$ & $_{-0.0418}^{+0.0522}$ \\
$\alpha$ & $-0.000957$ & $_{-0.00184}^{+0.00187}$ & $_{-0.00406}^{+0.00345}$ & $_{-0.00675}^{+0.00523}$ \\
$B_s$ & $0.217$ & $_{-0.0168}^{+0.0167}$ & $_{- 0.0308}^{+0.0359}$ & $_{-0.0457}^{+0.0573}$ \\
$c^2_{s,eff}$ & $0.00155$ & $_{-0.00155}^{+0.000319}$ & $_{-0.00155}^{+0.00241}$ & $_{-0.00155}^{+0.00493}$ \\
$n_s$ & $0.999$ & $_{-0.0227}^{+0.0230}$ & $_{-0.0393}^{+0.0509}$ & $_{-0.0548}^{+0.0811}$ \\
$\log[10^{10} A_s]$ & $3.0721$ & $_{-0.0362}^{+0.0370}$ & $_{-0.0713}^{+0.0730}$ & $_{-0.112}^{+0.111}$ \\
\hline
$\Omega_d$ & $0.956$ & $_{-0.00252}^{+0.00249}$ & $_{-0.00530}^{+0.004673}$ & $_{-0.00840}^{+0.00679}$ \\
$Age/Gyr$ & $13.723$ & $_{-0.145}^{+0.147}$ & $_{-0.280}^{+0.295}$ & $_{-0.435}^{+0.446}$ \\
$\Omega_b$ & $0.0439$ & $_{-0.00249}^{+0.00252}$ & $_{-0.00467}^{+0.00530}$ & $_{-0.00679}^{+0.00841}$ \\
$z_{re}$ & $10.924$ & $_{-1.242}^{+1.260}$ & $_{-2.410}^{+2.498}$ & $_{-3.824}^{+3.873}$ \\
$H_0$ & $72.189$ & $_{-1.816}^{+1.824}$ & $_{-3.631}^{+3.626}$ & $_{- 5.311}^{+5.537}$ \\
\hline\hline
\end{tabular}
\caption{The cosmological model and derived model parameters space with $1,2,3\sigma$ regions where SN Union2.1+BAO+CMB data sets are used.}\label{tab:results}
\end{table}
\end{center}

\begin{widetext}
\begin{center}
\begin{figure}[htb]
\includegraphics[width=18cm]{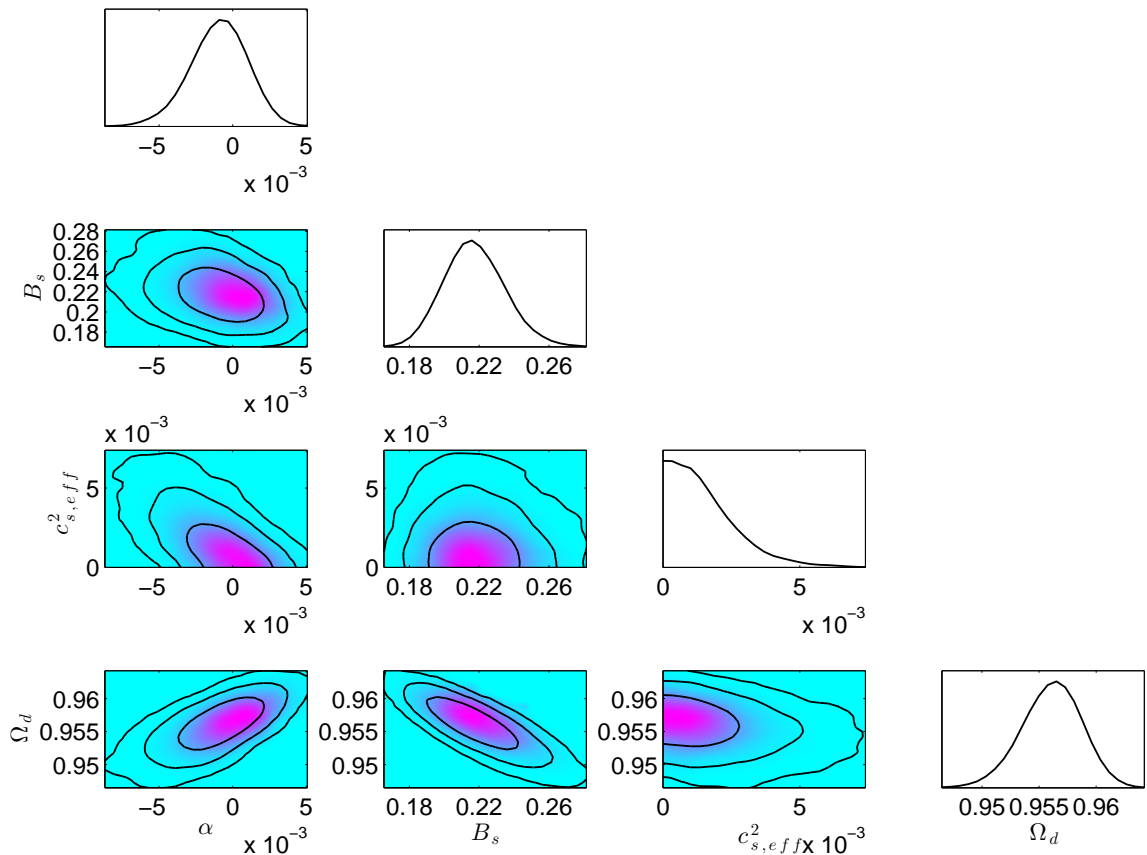}
\caption{The $1-D$ likelihood and $2-D$ contour plots for the model parameters with $1, 2, 3\sigma$ regions.}\label{fig:contour}
\end{figure}
\end{center}
\end{widetext}

\begin{center}
\begin{figure}[htb]
\includegraphics[width=9.cm]{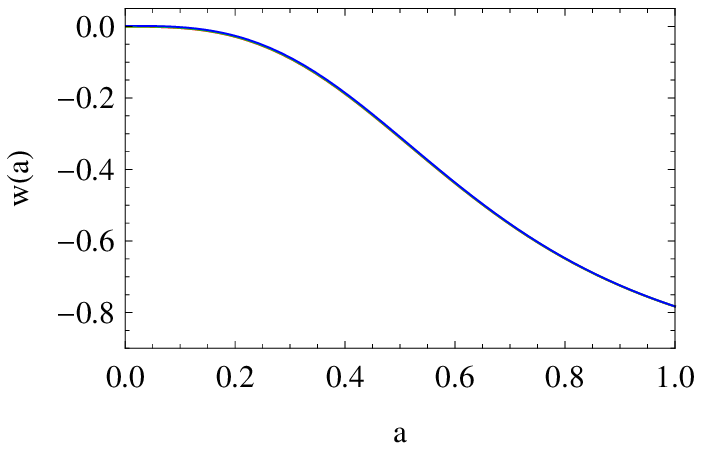}
\caption{The evolutions of the EoS $w_d(a)$ for UDF with respect to the scale factor $a$, where the mean values of the relevant model parameters are adopted.}\label{fig:eos}
\end{figure}
\end{center}

\begin{center}
\begin{figure}[tbh]
\includegraphics[width=9.cm]{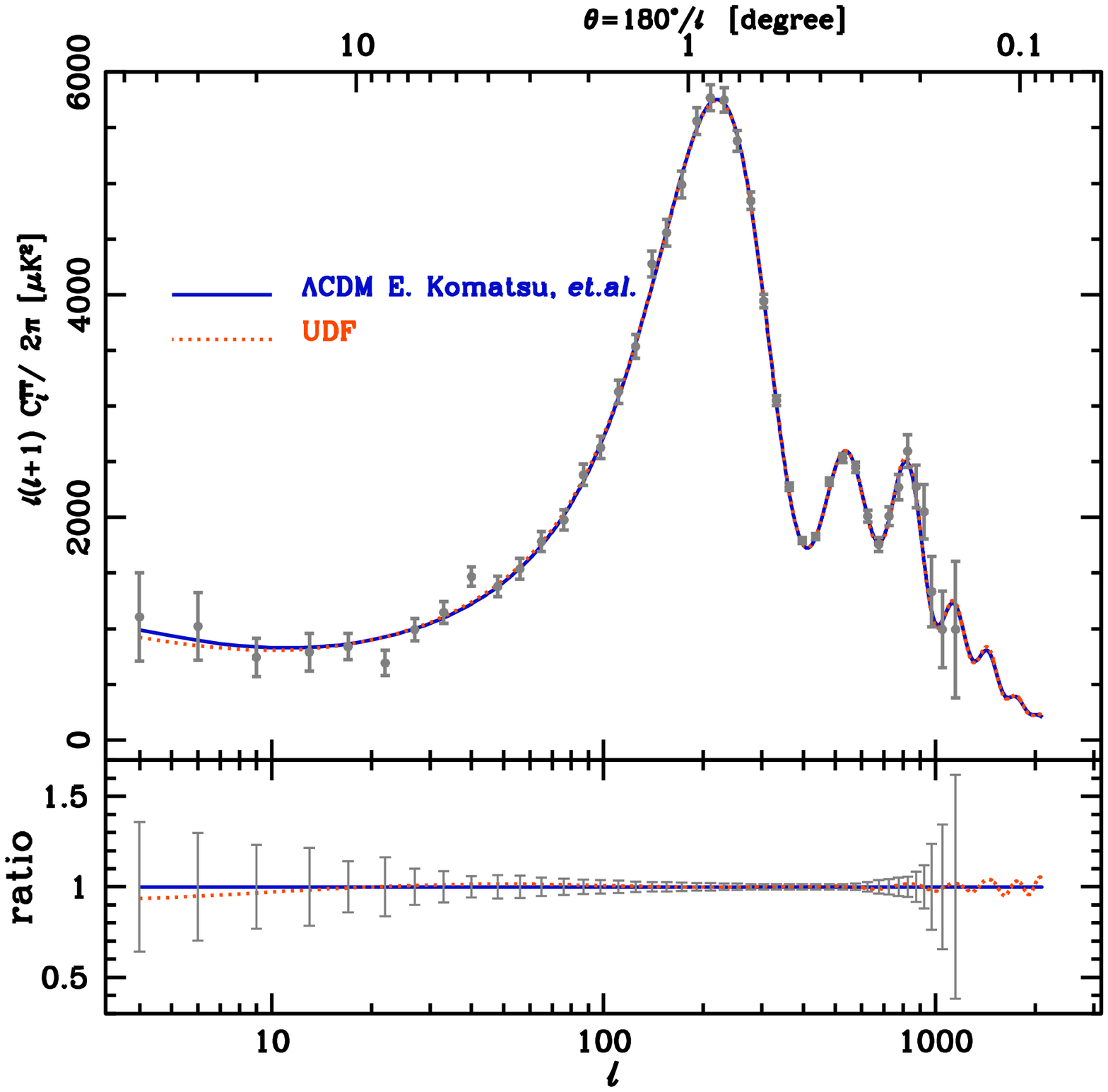}
\caption{The CMB $C^{TT}_l$ power spectrum v.s. multiple moment $l$, where the grey dots with error
bars denote the observed data with their corresponding uncertainties from WMAP $7$-year results, the red dashed lines are for the UDF model with mean values as shown in Table \ref{tab:results}, the blue solid lines are for $\Lambda$CDM model with mean values taken from \cite{ref:wmap7} with WMAP+BAO+$H_0$ constraint results. The bottom panel shows the ratios to $\Lambda$CDM model.}\label{fig:mean}
\end{figure}
\end{center}

The most relevant and interesting model parameters to the UDF are the constant adiabatic sound speed $c^2_{s,ad}=\alpha$ and the effective sound speed $c^2_{s,eff}$. They equal to each other when the entropy perturbation vanishes. The results show that the currently available data sets from SN, CMB and BAO can give a tight constraint to the model parameter space and favor a model with $c^2_{s,eff}= 0.00155_{-    0.00155}^{+    0.000319}$ in $1\sigma$ region. When the entropy perturbation is included, a negative adiabatic sound speed is favored which is different from that of the pure adiabatic case. By using the obtained mean values, we plotted the evolutions of EoS for the UDF with respect to the scale factor $a$ in Figure \ref{fig:eos}. From the EoS figure, one can read off that the UDF behaves like cold dark matter at the earlier epoch and like dark energy at the later epoch. Then the small perturbations of UDF can grow into the large scale structure of our Universe. We also show the $C^{TT}_{l}$ power spectra for $\Lambda$CDM model and observed data points in Figure \ref{fig:mean} where the mean values of relevant model parameters are adopted. It implies that current cosmic observational data points can not discriminate $\Lambda$CDM model from the UDF model. 

\begin{center}
\begin{figure}[tbh]
\includegraphics[width=9cm]{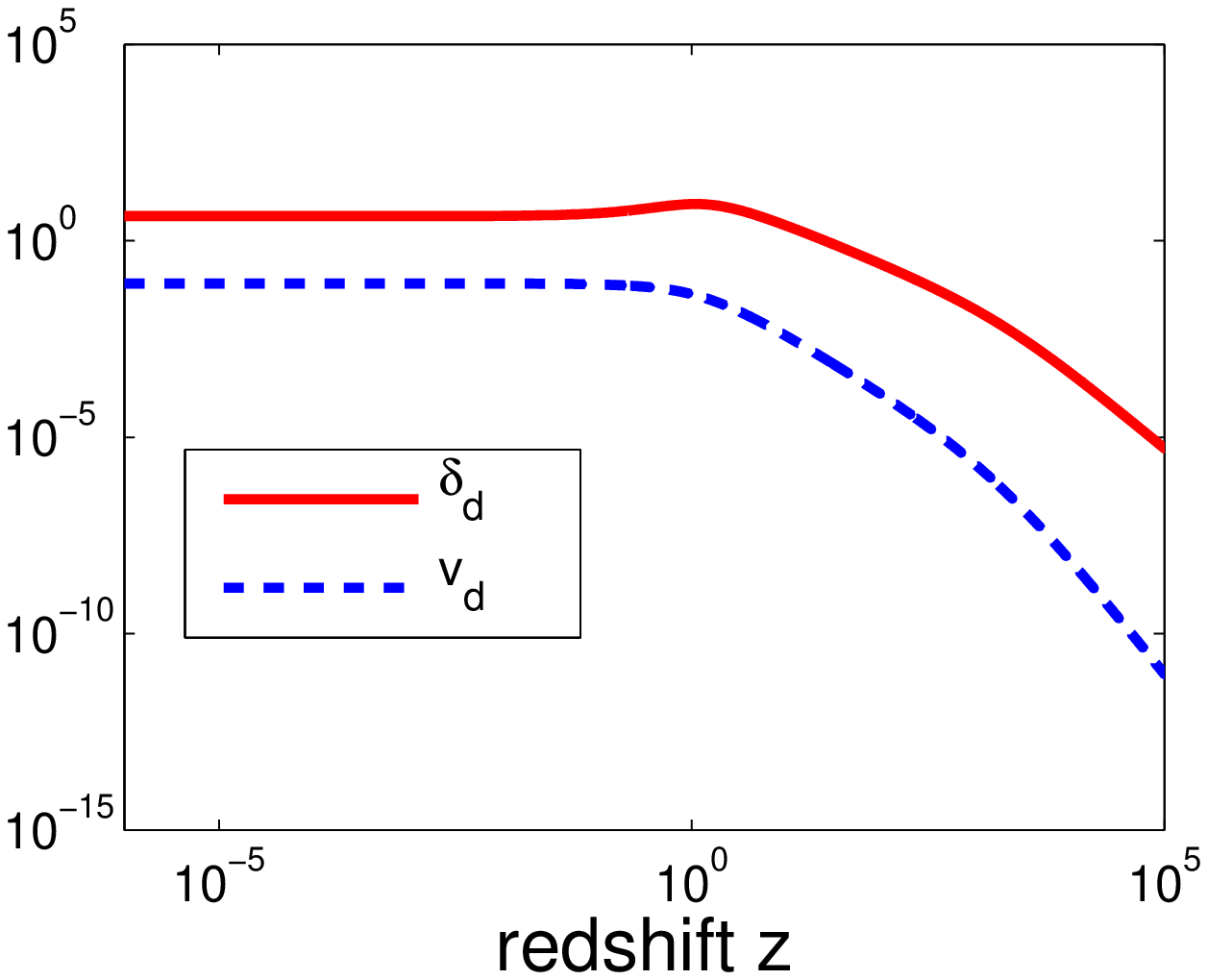}
\caption{Evolution of $\delta_d$ (thin) and $v_d$ (dashed) with respect to the redshift $z$ where the mean values of model parameters listed in Table \ref{tab:results} are adopted. Here $k=10^{-3}\text{Mpc}^{-1}$.}\label{fig:deperz}
\end{figure}
\end{center}

We also show the evolutions of $\delta_d$ and $v_d$ of the UDF with respect to the redshift $z$ on the scale $k=10^{-3}\text{Mpc}^{-1}$ in Figure \ref{fig:deperz}. One can easily see the growth of the UDF perturbations with the evolution of our Universe.

\section{Summary} \label{ref:conclusion} 

In this paper, we continued to study a unified dark fluid model with a constant adiabatic sound speed but with the entropic perturbations. As comparisons to the previous work \cite{ref:alphaentropy}, we took the effective sound speed as a free model parameter to characterize the micro-scale properties in additional to its EoS instead of fixing it to a specific value by hand. And then we try to use the currently available cosmic observations which include SN Union2.1, BAO and full information of CMB to determine the properties of this UDF model via the MCMC method. A tight constraint was obtained as shown in Table \ref{tab:results}. We also show the evolutions of the EoS of the UDF and its perturbations with the evolution of our Universe. The analysis tells us that the UDF behaves like cold dark matter at early epoch and like dark energy at late time. One can also see the growth of the perturbations of the UDF with the expansion of our Universe in Figure \ref{fig:deperz}. It shows the possibility for a small UDF perturbation to grow into a large scale structure of the Universe. We expect our study can shed light on the understanding of the dark side of our Universe.

\acknowledgements{The author thanks an anonymous referee for helpful improvement of this paper. L. Xu's work is supported in part by NSFC under the Grants No. 11275035 and "the Fundamental Research Funds for the Central Universities" under the Grants No. DUT13LK01.}


\begin{thebibliography}{*}

\bibitem{ref:darkdeneracy} M. Kunz., Phys. Rev. D 80, 123001 (2009); W. Hu and D. J. Eisenstein, Phys. Rev. D 59, 083509(1999); C. Rubano and P. Scudellaro, Gen. Relativ. Gravit. 34, 1931 (2002); I. Wasserman, Phys. Rev. D 66, 123511 (2002); A. R. Liddle and L. A. Urena-Lopez, Phys. Rev. Lett. 97, 161301 (2006); M. Kunz, A. R. Liddle, D. Parkinson, and C. Gao, Phys. Rev. D 80, 083533 (2009); A. Avile«s and J. L. Cervantes-Cota, Phys. Rev. D 83, 023510 (2011); L. M. Reyes, J. E. Madriz Aguilar, L.A. Urena-Lopez, Phys. Rev. D 84, 027503 (2011); A. Aviles, J. L. Cervantes-Cota, Phys. Rev. D 84, 083515 (2011); S. Camera, D. Bertacca, A. Diaferio, N. Bartolo, S. Matarrese, Mon. Not. Roy. Astron. Soc. 399,1995(2009) [arXiv:0902.4204]; S. Camera, T. D. Kitching, A. F. Heavens, D. Bertacca, A. Diaferio, Mon. Not. Roy. Astron. Soc. 415, 399(2011) [arXiv:1002.4740]; O. Luongo, H. Quevedo, [arXiv:1104.4758]; O. Luongo, H. Quevedo, DOI: 10.1007/s10509-011-0937-x; 
E. A. Lim, I. Sawicki, A. Vikman, JCAP 1005, 012 (2010); S. Capozziello, S. Nojiri, S.D. Odintsov, Phys. Lett. B632, 597(2006); S. Capozziello, V.F. Cardone, E. Elizalde, S. Nojiri,  S.D. Odintsov, Phys. Rev. D73, 043512(2006); K. Bamba, S. Capozziello, S. Nojiri, S. D. Odintsov, Astrophys.Space Sci. 342, 155(2012); H. Velten, D. J. Schwarz, JCAP 1109, 016 (2011); W.S. Hipolito-Ricaldi, H.E.S. Velten, W. Zimdahl, Phys. Rev. D 82 063507(2010). 

\bibitem{ref:Bruni} K. N. Ananda and M. Bruni, Phys. Rev. D. 74, 023523
(2006), arXiv:astro-ph/0512224; A. Balbi, M. Bruni, C. Quercellini, Phys. Rev. D  76, 103519 (2007).  

 \bibitem{ref:darkdeneracyxu}  L. Xu, Y. Wang, H. Noh, Phys. Rev. D 85, 043003 (2012) [arXiv:1112.3701].


\bibitem{GCG} A.Y. Kamenshchik, U. Moschella and V. Pasquier, 2001 {\it Phys. Lett. B} {\bf 511} 265.

\bibitem{GCG-action} M. C. Bento, O. Bertolami and A. Sen, Phys. Rev. D 66, 043507 (2002).

\bibitem{GCGpapers} T. Barreiro, O. Bertolami and P. Torres,  2008 {\it Phys. Rev. D} \textbf{78} 043530;
 M. Makler, S.Q. Oliveira and I. Waga, 2003 {\it Phys. Lett. B} \textbf{555} 1;
 R. Bean, O. Dore, 2003 {\it Phys. Rev. D} \textbf{68}  023515;
 L. Amendola, L.F. Finelli, C. Burigana, D. Carturan, 2003  {\it J. Cosmol. Astropart. Phys.} \textbf{0307} 005;
 A. Dev, D. Jain, J.S. Alcaniz, 2004 {\it Astron. Astrophys.} \textbf{417} 847;
 J.B. Lu {\it et al},  2008 {\it Phys. Lett. B}  {\bf 662}, 87;
 O.F. Piattella, JCAP 1003:012,2010;
 V. Gorini, A.Y. Kamenshchik, U. Moschella, O.F. Piattella, A.A. Starobinsky, JCAP 0802:016,2008; M. C. Bento, O. Bertolami, and A. A. Sen, Phys. Rev. D 66, 043507 (2002); L. Xu, J. Lu, JCAP 1003, 025(2010); J. Lu, Y. Gui, L. Xu, Eur. Phys. J. C 63,349(2009); N. Liang, L. Xu, Z. H. Zhu, Astrono. \& Astrophy, 527, A11(2011); Z. Li, P. Wu, H. Yu, JCAP09,017(2009); P. Wu, H. Yu, Phys. Lett. B 644,16(2007); C. G. Park, J. c. Hwang, J. Park, H. Noh, Phys. Rev. D 81,063532(2010).
 
\bibitem{GCGdecomp}  M. C. Bento, O. Bertolami, and A. A. Sen, Phys. Rev.
D, 70, 083519 (2004). 

\bibitem{GCGxu} L. Xu, J. Lu, Y. Wang, Eur. Phys. J. C 72 1883 (2012).

\bibitem{ref:xuNUDF} L. Xu, arXiv:1210.5327 [astro-ph.CO]. 

\bibitem{ref:alphcdm} A. Balbi, M. Bruni, C. Quercellini, Phys. Rev. D  76, 103519 (2007).


 \bibitem{ref:cs0} A. Aviles, J. L. Cervantes-Cota, Phys. Rev. D 84, 083515 (2011); O. Luongo, H. Quevedo, [arXiv:1104.4758]; O. Luongo, H. Quevedo, DOI: 10.1007/s10509-011-0937-x. 
 
 \bibitem{ref:csvar} S. Camera, D. Bertacca, A. Diaferio, N. Bartolo, S. Matarrese, Mon. Not. Roy. Astron. Soc. 399,1995(2009) [arXiv:0902.4204]; S. Camera, T. D. Kitching, A. F. Heavens, D. Bertacca, A. Diaferio, Mon. Not. Roy. Astron. Soc. 415, 399(2011) [arXiv:1002.4740].
 
 \bibitem{ref:Riess98} A.G. Riess, {\it et al.}, Astron. J. 116 1009(1998) [astro-ph/9805201]. 

\bibitem{ref:Perlmuter99} S. Perlmutter, {\it et al.}, Astrophys. J. 517, 565(1999) [astro-ph/9812133].

 \bibitem{alphanegative} C. Quercellini, M. Bruni, and A. Balbi, Classical and Quantum Gravity 24, 5413 (2007), arXiv:0706.3667.

\bibitem{ref:stability} L.M.G. Beca, P.P. Avelino, Mon. Not. Roy. Astron. Soc. 376,1169(2007); P. P. Avelino, L. M. G. Be\c{c}a, C. J. A. P. Martins, Phys. Rev. D 77, 063515 (2008).
 
\bibitem{ref:Hu98} 	W. Hu, Astrophys. J. 506, 485(1998).
 
\bibitem{ref:alphaentropy} D. Pietrobon, A. Balbi, M. Bruni, C. Quercellini, Phys. Rev. D 78, 083510(2008), arXiv:0807.5077 [astro-ph].


\bibitem{ref:MB} C.-P Ma and E. Bertschinger, Astrophys. J. 455, 7 (1995).

\bibitem{ref:Hwang} J. Hwang, H. Noh, Phys. Rev. D 65,023512(2001).

\bibitem{ref:WangXu} Y. Wang, L. Xu, Y. Gui, Phys. Rev. D 84, 063513(2011).


\bibitem{ref:k-essence} C. Armendariz-Picon, V. Mukhanov, P. J. Steinhardt, Phys. Rev. D63:103510(2001).

\bibitem{ref:soundspeed} J. Valiviita, E. Majerotto, R. Maartens, JCAP 020, 0807(2008); L. Xu, Y. Wang, H. Noh, 
Phys. Rev. D. 84, 123004(2011).

\bibitem{ref:MCMC} http://cosmologist.info/cosmomc/; A. Lewis and S. Bridle, Phys. Rev. D 66, 103511 (2002).

\bibitem{ref:CAMB}  http://camb.info/.

\bibitem{ref:bbn} S. Burles, K. M. Nollett, and M. S. Turner, Astrophys. J.
552, L1 (2001).

\bibitem{ref:hubble} A. G. Riess et al., Astrophys. J. 699, 539 (2009).


\bibitem{ref:wmap7} E. Komatsu et al., Astrophys. J. Suppl. Ser. 192, 18 (2011); http://lambda.gsfc.nasa.gov/product/map/current/.

\bibitem{ref:BAO} W. J. Percival et al., Mon. Not. R. Astron. Soc. 401, 2148
(2010).

\bibitem{ref:Union21} N. Suzuki, et al. (Supernova Cosmology Project
Collaboration), arXiv:1105.3470 [astro-ph.CO], http://supernova.lbl.gov/Union/.

\bibitem{ref:Xu} L. Xu, Y. Wang, JCAP, 06, 002(2010); L. Xu, Y. Wang, Phys. Rev. D 82, 043503 (2010).


\end{thebibliography}
\end{document}